\documentclass[letterpaper,titlepage,11pt]{article}

\pdfoutput=1

\usepackage{amssymb,amsmath,amsfonts}
\usepackage{epsfig}
\usepackage{ccaption}
\usepackage{graphicx}

\usepackage[
      colorlinks=true,
      linkcolor=blue,
      urlcolor=blue,
      filecolor=black,
      citecolor=red,
      pdfstartview=FitV,
      pdftitle={},
        pdfauthor={Roberto Emparan, Kentaro Tanabe},
        pdfsubject={},
        pdfkeywords={},
        pdfpagemode=None,
        bookmarksopen=true
      ]{hyperref}

\def\clock{{\count0=\time
           \divide\count0 60
           \ifnum\count0<10 0\fi\the\count0
           \multiply\count0 -60 \advance\count0 \time
           :\ifnum\count0<10 0\fi \the\count0
         }}
\newcommand{\timestamp}{{\small\vbox{\hbox{\tt\jobname.tex}
\hbox{\the\day/\the\month/\the\year, \clock}}}}

\newcommand{\ie}{{\it i.e.,\,}}
\newcommand{\eg}{{\it e.g.,\,}}

\newcommand{\lp}{\left(}
\newcommand{\rp}{\right)}

\newcommand{\beq}{\begin{equation}}
\newcommand{\eeq}{\end{equation}}
\newcommand{\bea}{\begin{eqnarray}}
\newcommand{\eea}{\end{eqnarray}}
\newcommand{\beqa}{\begin{eqnarray}}
\newcommand{\eeqa}{\end{eqnarray}}

\newcommand{\R}{\mathbb{R}}

\newcommand{\sR}{\mathsf{R}}

\numberwithin{equation}{section}

\begin{document}

\begin{titlepage}
\leftline{}
\vskip 1cm
\centerline{\LARGE \bf Holographic superconductivity} 
\medskip
\centerline{\LARGE \bf in the large $D$ expansion} 

\vskip 1.6cm
\centerline{\bf Roberto Emparan$^{a,b}$, Kentaro Tanabe$^{b}$}
\vskip 0.5cm

\centerline{\sl $^{a}$Instituci\'o Catalana de Recerca i Estudis
Avan\c cats (ICREA)}
\centerline{\sl Passeig Llu\'{\i}s Companys 23, E-08010 Barcelona, Spain}
\smallskip
\centerline{\sl $^{b}$Departament de F{\'\i}sica Fonamental, Institut de
Ci\`encies del Cosmos,}
\centerline{\sl  Universitat de
Barcelona, Mart\'{\i} i Franqu\`es 1, E-08028 Barcelona, Spain}

\vskip 0.5cm
\centerline{\small\tt  emparan@ub.edu,\, ktanabe@ffn.ub.es}

\vskip 1.6cm
\centerline{\bf Abstract} \vskip 0.2cm \noindent

We study holographic superconductivity by expanding the equations in the inverse of the number of spacetime dimensions $D$. We obtain an analytic expression for the critical temperature as a function of the conformal dimension of the condensate operator. Its accuracy for $3+1$-dimensional superconductors is better than $15\%$. The analysis reveals a simple, and quantitative, explanation for the onset of the superconducting instability, as well as universal features of holographic superconductivity in the large $D$ limit. In particular, this allows to easily compute the effects of backreaction on the critical temperature.

\end{titlepage}
\pagestyle{empty}
\small
\normalsize
\newpage
\pagestyle{plain}
\setcounter{page}{1}


\newpage

\section{Introduction}

The study of the limit where the number of spatial dimensions grows to infinity is a strategy that is well known in statistical mechanics --- lattices of infinite coordination render mean-field theory exact \cite{Georges:1996zz} --- but is seldom used in continuum field theories. This neglect seems reasonable in quantum field theories, since ultraviolet divergences grow infinitely unruly in this limit, but most classical field theories, and in particular General Relativity, remain well defined in higher dimensions. For them, the spacetime dimension $D$ can be sensibly regarded as a parameter to be varied. Refs.~\cite{Emparan:2013moa,Emparan:2013xia} (see also previous work in \cite{Asnin:2007rw}) have identified several aspects that make the expansion in $1/D$ a useful approach to General Relativity. In particular, the large field gradients near a black hole horizon induce a parametric separation between a `far zone', with trivial interactions, and a `near zone', with non-trivial but universal string-type near-horizon geometries.\footnote{Previous partial observations were made in \cite{Soda:1993xc,Grumiller:2002nm}.} In ref.~\cite{Emparan:2013moa} we have shown how this allows to solve in analytic form several black hole phenomena, with results that remain accurate even at relatively low values of $D$.

Here we shall apply the large $D$ expansion to the phenomenon of holographic superconductivity \cite{Hartnoll:2008vx,Hartnoll:2008kx}. One purpose is to demonstrate the technique in a non-trivial AdS/CFT setting. We find that analytical results can be readily obtained which are quantitatively good for `realistic' values of $D$. Moreover, the study reveals a simple, universal structure in holographic superconductors in the large $D$ limit. 

In the large $D$ analysis the onset of the superconducting instability is qualitatively transparent and quantitatively easy to determine. It occurs when a `near-zone bound' on the field parameters, analogous to the Breitenlohner-Freedman (BF) bound, is violated in the near-horizon region. Meanwhile, the field in the far region, which is essentially the AdS geometry, satisfies the corresponding `far-zone bound' and therefore the complete field configuration is stable.\footnote{A similar argument \cite{Denef:2009tp,Horowitz:2009ij,Hartnoll:2011fn} explains a condensation instability at very low temperatures near extremality (see sec.~\ref{sec:backr}). Our bound is different --- more general and valid arbitrarily away from extremality.} In contrast, if the far region were asymptotically flat the configuration would be tachyonic everywhere, so the non-trivial scalar hair would not form. 
These BF-type bounds are easily evaluated, and they directly give quantitative information on the critical points of the superconducting transition. 
Furthermore, the features that determine these points are shared by large classes of black holes and black branes ---\eg\ with or without a cosmological constant, with or without charge \cite{Emparan:2013xia} --- and so the phenomenon exhibits universal properties that relate them all. We have found that this makes surprisingly easy to include backreaction effects.

In the next section we present our simplest example, solve the equations in detail, and test the accuracy of our analytical result against numerical calculations. In sec.~\ref{sec:simple} we distil the main components of this analysis into a simpler and more intuitive understanding of how superconducting condensates form at large $D$. In sec.~\ref{sec:backr} we show how this allows to incorporate gauge field backreaction almost effortlessly. We then conclude in sec.~\ref{sec:concl}.

\section{Large $D$ holographic superconductivity}
\label{sec:testmodel}

\subsection{Set up}

We begin with the simplest model for a holographic superconductor \cite{Hartnoll:2008vx}: a massive scalar $\Psi$ charged under an Abelian gauge field $A_\mu$ in AdS$_{n+1}$, with action
\beq
I=-\int d^{n+1}x \sqrt{-g}\lp \frac14 F^2 +|\nabla\Psi-i A\Psi|^2+m^2|\Psi|^2\rp\,.
\eeq
Instead of the number of bulk spacetime dimensions $D$ we use $n=D-1$, \ie\ the spacetime dimensionality of the dual boundary theory. Following \cite{Hartnoll:2008vx} we consider the regime of large charge where the backreaction of these fields on the geometry is negligible. The background spacetime is the AdS black brane
\beq\label{adsbb}
ds^2=-r^2 h(r)dt^2+\frac{dr^2}{r^2 h(r)}+r^2 d\mathbf{x}^2_{n-1}\,,
\eeq
\beq
h(r)= 1-\frac{r_0^n}{r^n}\,,
\eeq
with temperature
\beq
T=\frac{n r_0}{4\pi}\,.
\eeq
We have set the AdS radius to one. The field equation for $\Psi$ requires that its phase be constant and we set it to be real, $\Psi=\psi(r)$. For the gauge field we take $A=\phi(r)dt$. The field equations are
\beq\label{scaleq}
\psi^{''}+\lp \frac{h'}{h}+\frac{n+1}{r}\rp\psi'+\lp\frac{\phi^2}{r^4 h^2}-\frac{m^2}{r^2 h}\rp\psi=0\,,
\eeq
and
\beq\label{gaugeeq}
\phi^{''}+\frac{n-1}{r}\phi'-\frac{2\psi^2}{r^2 h}\phi=0\,.
\eeq
Regularity at the horizon requires that $\psi$ remains finite and $\phi$ vanishes at $r=r_0$.
At infinity we demand that their leading asymptotic behavior be
\beqa
\psi &=&\frac{\psi_+}{r^\lambda}+\dots\,,\\
\phi&=&\mu -\frac{\rho}{r^{n-2}}+\dots\,,
\eeqa
where
\beq\label{lambval}
\lambda=\frac{n}{2}\lp 1+\sqrt{1+4\hat m^2}\rp\,, \qquad \hat m= \frac{m}{n}\,.
\eeq
With this choice of asymptotic fall-off for the scalar field,  $\lambda$
is the conformal dimension of the operator dual to $\psi$ in the boundary theory, whose expectation value, \ie\ the order parameter of the condensate, is given by $\psi_+$. We will not consider the alternative quantization that selects the other independent solution, $\sim r^{\lambda-n}$, which is normalizable when $m^2<-(n+1)^2/4+1$.

In order to take the large $n$ limit of the system of equations \eqref{scaleq}, \eqref{gaugeeq} we need to spell out how the different terms in them scale with $n$.
A crucial property of the large $n$ limit of the gravitational field of a black hole is that its gradients become large, of order $n$, in a small region of extent $\sim r_0/n$ outside the horizon. We expect the condensation of the scalar field to occur in this near-horizon region. There, generically, a radial derivative will be $\propto n$. 

Focus first on the scalar equation \eqref{scaleq}. We want the gauge potential term $\propto \phi^2$ to survive in the large $n$ limit, as it is the one that drives the condensation of the scalar.
Thus it must be $\phi\sim n$. 
The mass term will remain significant at large $n$ if $m\sim n$, so we take $\hat m=O(n^0)$.\footnote{Hats are used to refer to variables and quantities rescaled by factors of $n$, which naturally belong in the near-horizon region.} Hence,
all masses that are $|m|< O(n)$ are essentially zero (to leading order) at large $n$. We will consider masses strictly above the BF bound $ m^2_{BF}=-(n+1)^2/4$, \ie\
\beq\label{bfbound}
\hat m^2 >-\frac{1}{4}+O(1/n)\,.
\eeq

Now consider the gauge field equation \eqref{gaugeeq} in the near-region where radial derivatives are $\propto n$.
If $\psi \sim n$ the system remains a non-linear set of field equations which is difficult to solve. Instead, if $\psi$ is comparatively small, $\psi= O(n^a)$, $a<1$, then its effects on the gauge field are negligible. Although we do not know a priori what the size (in $n$) of the condensate will be in general, it will be small enough if the system is sufficiently close to the critical point. 

\subsection{Gauge field}

Under the last assumption, we can neglect the scalar in the gauge field equation in the near zone.
On the other hand, in the far zone,  $r-r_0\gg r_0/n$, the effect of the scalar on the gauge field equations is also negligible since it falls off exponentially fast in $n$ compared to other terms. Then, the gauge field equation everywhere outside the horizon becomes
\beq\label{fargaugeeq}
\phi^{''}+\frac{n-1}{r}\phi'=0\,,
\eeq
which we solve as
\beq\label{phisol}
\phi=n\hat\mu r_0 \lp 1-\lp\frac{r_0}{r}\rp^{n-2}\rp
\eeq
in terms of a dimensionless parameter $\hat\mu$  (of order one in $n$) which determines both the chemical potential $\mu=n\hat\mu r_0$ and the charge density
\beq
\rho=n\hat\mu r_0^{n-1}\,.
\eeq
Owing to a scaling symmetry of the system, neither $\rho$ nor the temperature $T$ have absolute meaning, instead the state is characterized by the dimensionless ratio\footnote{We could equally well use $T/\mu=1/(4\pi\hat\mu)$ but it is common to employ \eqref{Trhomu}.}
\beq\label{Trhomu}
\frac{T}{\rho^{1/(n-1)}}=\frac{n}{4\pi}(n\hat\mu)^{-\frac1{n-1}}\,.
\eeq
We will take $\hat\mu$ as the control parameter. For fixed $\rho$ we regard a large (small) value of $\hat\mu$ as a small (large) temperature.

\subsection{Scalar field}

At large $n$ we can identify the solution \eqref{phisol} with $\phi =n\hat\mu r_0 h$. Inserting this in \eqref{scaleq} we obtain
\beqa\label{scaleq1}
\psi^{''}+\lp \frac{h'}{h}+\frac{n+1}{r}\rp\psi'+n^2\lp \frac{r_0^2}{r^4}\,\hat\mu^2-\frac{\hat m^2}{r^2 h}\rp\psi=0\,.
\eeqa
We shall solve this linear equation for $\psi$ in the manner explained in \cite{Emparan:2013moa,Asnin:2007rw}: we first solve the equations in the near-horizon zone $r-r_0\ll r_0$, requiring regularity at the horizon; next, we solve them in the far-zone $r-r_0\gg r_0/n$, imposing absence of sources for the scalar. Finally, we match these two solutions in their common overlap zone $r_0/n\ll r-r_0\ll r_0$. Performing this matching is equivalent to solving the full boundary value problem, which will admit a solution only for certain values of $\hat\mu$.

The study of eq.~\eqref{scaleq1} in this section is quite thorough, which is necessary in order to settle details of the computation. However, it contains a simple core argument that we extract in sec.~\ref{sec:simple}. Some readers may prefer to read that section first and then return here for a more meticulous view.

\subsubsection{Near-zone}
\label{sec:nearsol}

In the region $r-r_0\ll r_0$ we introduce a new radial coordinate
\beq\label{sR}
\sR =\lp\frac{r}{r_0}\rp^n
\eeq
that remains finite as $n\to\infty$. In this limit
\beq\label{nearR}
\frac{d}{dr}\to \frac{n}{r_0}\sR\frac{d}{d\sR}\,,\qquad h\to  1-\sR^{-1}\,,\qquad
h'\to \frac{n}{r_0\sR}\,.
\eeq

\paragraph{Near-zone BF-type bound.}

The near-horizon geometry that arises from \eqref{adsbb} is
\beq\label{neards2}
ds^2|_\mathrm{nh}=\frac{r_0^2}{n^2}\lp -(1-\sR^{-1}) d\hat t^2+\frac{d\sR^2}{\sR(\sR-1)} +d\hat{\bf x}^2_{n-1}\rp
\eeq
where $(\hat t, \hat{\bf x})=n r_0 (t,\mathbf{x})$.
The change $\sR=\cosh^2\rho$ brings the $(\hat t,\rho)$ part of the metric into a more familiar form of the two-dimensional string theory black hole \cite{2dbh} ($r_0/n$ plays the role of the string length). The size along the $\R^{n-1}$ directions of the brane plays the role of the dilaton \cite{Emparan:2013xia}. 

The asymptotic form as $\sR\to\infty$ of the non-trivial two-dimensional part of the metric in \eqref{neards2} is 
\beq\label{lindil}
ds^2_{2d}\to  -d\hat t^2+\frac{d\sR^2}{\sR^2}\,.
\eeq
It was argued in \cite{Emparan:2013xia} that this is the generic near-horizon asymptopia --- \ie\ the `overlap zone' --- of many families of black holes at large $D$.\footnote{The reason for this universality is that this linear dilaton background appears in the large $D$ limit of Minkowski space \cite{Emparan:2013xia}. For a non-vacuum (but non-hairy) black hole, the matter stress-energy tensor dies away at large $\sR$ sufficiently fast to become negligible and yield the linear dilaton vacuum. A cosmological radius that is $O(n)$ compared to the near-horizon region size only effects a change in the 2D black hole mass.}

Consider a scalar field $\psi$ that in the asymptotic region \eqref{lindil} satisfies the static field equation with (effective) mass $\hat m_\mathrm{eff}$, namely,
\beq\label{overlappsi}
\frac{d}{d\sR}\lp\sR^2\, \frac{d\psi}{d\sR}\rp -\hat m_\mathrm{eff}^2\psi=0\,.
\eeq
Then in this region the solution must be of the form
\beq\label{asympsol}
\psi = A_+ \sR^{-\hat\lambda_+}+A_-\sR^{-\hat\lambda_-} 
\eeq
with
\beq\label{hlambda}
\hat\lambda_\pm=\frac12 \pm \sqrt{\frac14 +\hat m_\mathrm{eff}^2} 
\eeq
and constant $A_\pm$. A straightforward analogue of the derivation of the BF stability bound in AdS \cite{BFbound} shows that fields with
\beq\label{nbfviol}
\hat m_\mathrm{eff}^2< -\frac14
\eeq
have tachyonic (unstable) behavior.

\paragraph{Near-horizon solution.}

With \eqref{sR}, \eqref{nearR}, the scalar equation \eqref{scaleq1} takes the form\footnote{Making $\sR\to 1-\sR$ this becomes the same as the massless scalar equation in Schwarzschild in \cite{Emparan:2013moa}.} 
\beq\label{nearsolph}
\frac{d}{d\sR}\lp \sR(\sR-1)\frac{d}{d\sR}\psi\rp +\hat\mu^2\lp 1-\sR^{-1}\rp \psi-\hat m^2\psi=0\,.
\eeq
Comparing to \eqref{overlappsi} we see that in the asymptotic region $\sR\gg 1$ the field has an effective mass
\beq
\hat m_\mathrm{eff}^2 =\hat m^2 -\hat\mu^2\,,
\eeq
so the gauge field tends to make the scalar less stable. It will undergo a tachyonic instability when \eqref{nbfviol} holds, \ie\ when $\hat\mu^2 > \hat\mu_c^2$ with
\beq\label{muc}
\hat\mu_c^2=  \hat m^2 +\frac14\,.
\eeq
We will presently see that this is indeed the superconducting instability leading to a scalar condensate in the complete solution. Note that this near-horizon instability can exist while complying with the far-zone AdS BF bound \eqref{bfbound}.

We can find the exact solution of \eqref{nearsolph} at all $\sR$ in terms of hypergeometric functions, and if we
demand regularity at the horizon $\sR=1$ the solution is
\beq\label{nearsolps}
\psi=\sR^{i\hat\mu}\, {}_2F_1(i\hat\mu+\hat\lambda_+,i\hat\mu+\hat\lambda_-,1;1-\sR)\,,
\eeq
up to a multiplicative real integration constant. Here the indices \eqref{hlambda} take the form
\beq
\hat\lambda_\pm=\frac12 \pm \sqrt{\hat\mu_c^2-\hat\mu^2}\,.
\eeq
Even though the solution \eqref{nearsolps} involves complex numbers, one can readily check that it is a real function of $\sR$ for all real values of $\hat\mu$ and $\hat m^2$. At large $\sR$ it asymptotes to the form \eqref{asympsol} with coefficients $A_\pm$ given in appendix~\ref{app:neov}.
Thus the horizon boundary condition determines that, if the scalar field is non-zero, both asymptotic components in \eqref{asympsol} must be present.

\subsubsection{Far-zone}

In the far zone we neglect terms that fall off exponentially fast in $n$. Then the scalar field equation becomes,
after substituting the solution for $\phi$, 
\beq\label{farscaleq2}
\psi^{''}+\frac{n+1}{r}\psi'+\frac{n^2}{r^2}\lp\hat\mu^2\,\frac{r_0^2}{r^2} -\hat m^2\rp\psi=0\,,
\eeq
which is solved in terms of Bessel functions as
\beq\label{farsoln}
\psi= r^{-n/2} J_{n-2\lambda}\lp\frac{n\hat\mu r_0}{r}\rp\,,
\eeq
up to a constant amplitude factor. The conformal dimension $\lambda$ of the dual operator, \eqref{lambval}, can be written as
\beq\label{lambdamuc}
\lambda=\frac{n}2+n\hat\mu_c\,,
\eeq
with $\hat\mu_c$ defined in \eqref{muc}.

In order to study this solution in the overlap zone where $r_0/n\ll r-r_0\ll r_0$, \ie\ $1\ll \ln\sR\ll n$, we take
\beq\label{overrR}
\frac{r}{r_0}=1+\frac{1}{n}\ln\sR+O\lp n^{-2}\rp
\eeq
and we apply the Debye expansion of Bessel functions to the solution \eqref{farsoln}. We find that, up to $O(1/n)$ corrections,
\beq\label{farover1}
\psi\to\hat B_+\sR^{-\hat\lambda_+}\,, \qquad \mathrm{for~} \hat\mu<\hat\mu_c
\eeq
and 
\beq\label{farover2}
\psi\to\hat A_+\sR^{-\hat\lambda_+}
+\hat A_-\sR^{-\hat\lambda_-} \,, \qquad \mathrm{for~}  \hat\mu>\hat\mu_c\,.
\eeq
The explicit values of $\hat B_+$, $\hat A_\pm$ can be found in appendix~\ref{app:debye}.
The important point is that when $\hat\mu<\hat\mu_c$ the solution \eqref{farover1} only contains the component $\sR^{-\hat\lambda_+}$ but not $\sR^{-\hat\lambda_-}$. This is a consequence of having imposed the asymptotic behavior $\psi\sim r^{-\lambda}$, \ie\ not having a source for the scalar at the AdS boundary. Instead, when $\hat\mu>\hat\mu_c$  the field \eqref{farover2} has the two components. Therefore, only in the latter case can we match to the near-zone solution \eqref{nearover}. 

In other words, the BF-type bound $\hat\mu<\hat\mu_c$ marks a critical value of
\beq\label{critT0}
\left.\frac{T}{\rho^{1/(n-1)}}\right|_c=\frac{n}{4\pi}(n\hat\mu_c)^{-\frac1{n-1}}=\frac{n}{4\pi}\lp \lambda-\frac{n}2\rp^{-\frac1{n-1}}\,,
\eeq
above which there cannot exist a non-trivial scalar condensate outside the horizon.

This is a non-trivial result, but we can, and must, do better. We have not proven yet that the condensate is actually possible for $\hat\mu>\hat\mu_c$. In fact, near the critical point the indices $\hat\lambda_\pm$ become degenerate and the expansions that give \eqref{farover1}, \eqref{farover2} cease to be valid. As a consequence, the result \eqref{critT0} will be slightly modified.\footnote{This modification is analogous to computing the quantum-mechanical ground state energy in a potential well, which yields a slightly larger result than the classical energy at the minimum of the well. See sec.~\ref{sec:bound}.}

\subsection{Solution at the critical point}

When $\hat\mu-\hat\mu_c=O\lp n^{-2/3}\rp$ the expansions \eqref{farover1}, \eqref{farover2} break down. Then we write
\beq\label{closecr}
\hat\mu=\hat\mu_c+\lp\frac{\hat\mu_c}{2n^2}\rp^{1/3}\delta\hat\mu
\eeq
with $\delta\hat\mu=O(n^0)$, and the correct expansion of the far-region solution \eqref{farsoln} in the overlap zone is
\beq\label{farov}
\psi=\sR^{-1/2}\lp\frac{2}{n\hat\mu_c}\rp^{1/3}\lp \mathrm{Ai}(-\delta\hat\mu)
+{\rm Ai}' (-\delta\hat\mu) \lp\frac{2\hat\mu_c^2}{n}\rp^{1/3}\ln\sR + O(n^{-2/3})
\rp\,,
\eeq
in terms of the Airy function $\mathrm{Ai}$. In this same regime, the general near-region solution in the overlap zone, instead of \eqref{asympsol}, takes the form
\beq\label{nearov}
\psi = \sR^{-1/2}\lp A + B\ln\sR + O(1/\sR)\rp\,.
\eeq
We can readily obtain the constants $A$ and $B$ from the solution \eqref{nearsolps} that is regular at the horizon, see appendix~\ref{app:neov}, but we will not need their precise values. Only note that they are of the same order in $n$.

The matching of these near and far solutions has to be performed consistently at each order in $n$. In \eqref{farov}, the term $\propto \mathrm{Ai}(-\delta\hat\mu)$ is of higher order in $n$ than the $\ln\sR$ term. So, while we can readily match the coefficients of $\ln\sR$ terms in \eqref{farov} and \eqref{nearov}, we must require that
\beq\label{aicond}
\mathrm{Ai}(-\delta\hat\mu)=0\,,
\eeq
\ie\  non-trivial solutions for the scalar field are possible only when $\hat\mu$ is in correspondence with zeroes of the
 Airy function $\mathrm{Ai}(x)$. These are a discrete series of \textit{negative} zeroes $x=-a_1,-a_2,\dots$. Hence $\delta\hat\mu>0$, confirming that when $\hat\mu<\hat\mu_c$ the scalar field does not admit non-trivial solutions. For $\hat\mu>\hat\mu_c$ we find a discrete family of solutions with $\delta\hat\mu=a_k$.\footnote{Ref.~\cite{Hartnoll:2008vx} mentioned that a discrete family of solutions appeared possible, as we have found.} The first one, with
\beq\label{azero}
a_1=2.33811\,,
\eeq
is the only one we will consider in the following. 
The correct value of the critical parameter is then
\beq\label{mucrit}
\hat\mu_\mathrm{crit}=\hat\mu_c+\lp\frac{\hat\mu_c}{2n^2}\rp^{1/3}a_1
\eeq
and in terms of the physical parameters of the system, \eqref{Trhomu} and \eqref{lambdamuc}, the critical point is at
\beqa\label{critT1}
\left.\frac{T}{\rho^{1/(n-1)}}\right|_\mathrm{crit}
=\frac{n}{4\pi}\Biggl[ \lambda-\frac{n}2 +a_1\lp \frac{\lambda}2-\frac{n}4\rp^{1/3} \Biggr]^{-\frac1{n-1}}\,.
\eeqa
This is our main result. 

\begin{figure}[t]
\begin{tabular}{cc}
\includegraphics[width=.46\textwidth]{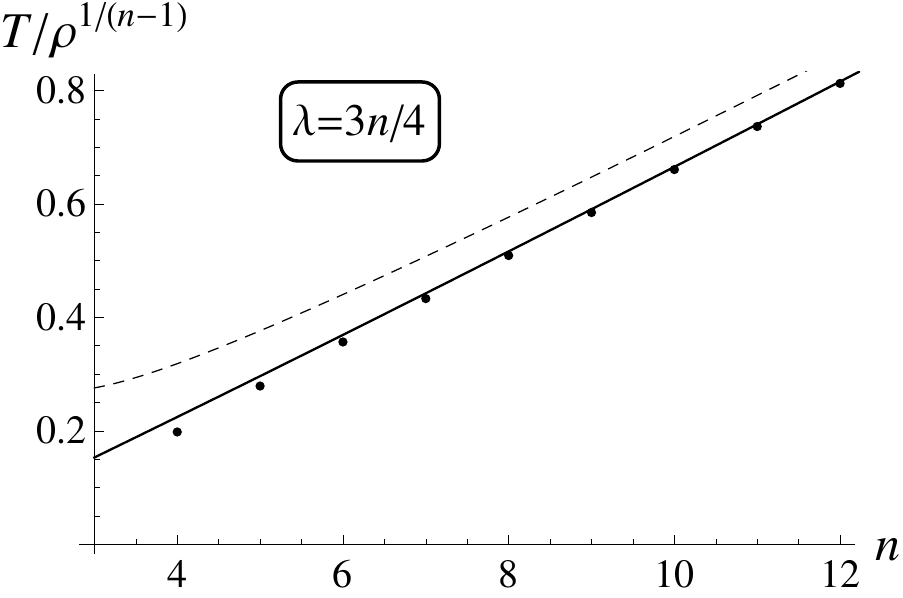}
\qquad
\includegraphics[width=.46\textwidth]{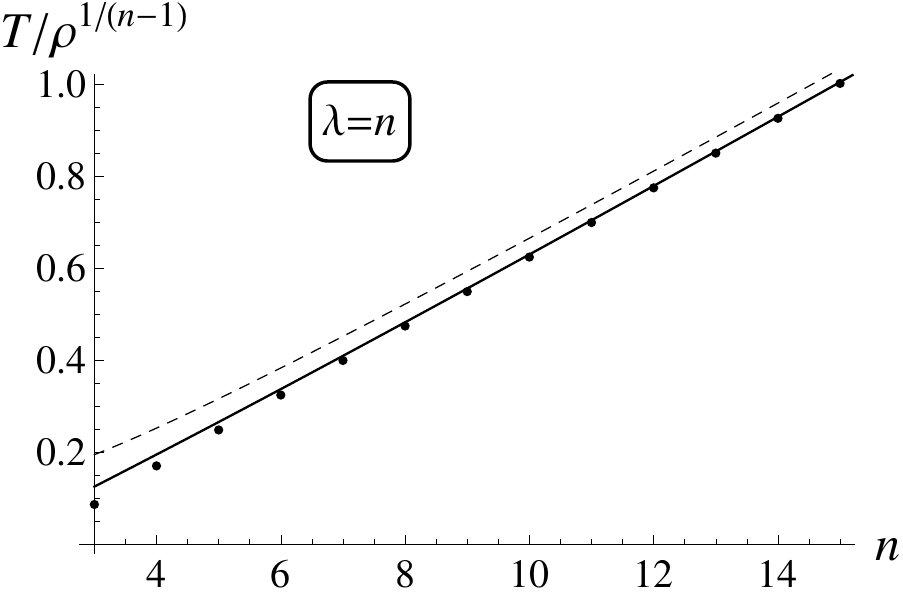}
\\{}\\
\includegraphics[width=.46\textwidth]{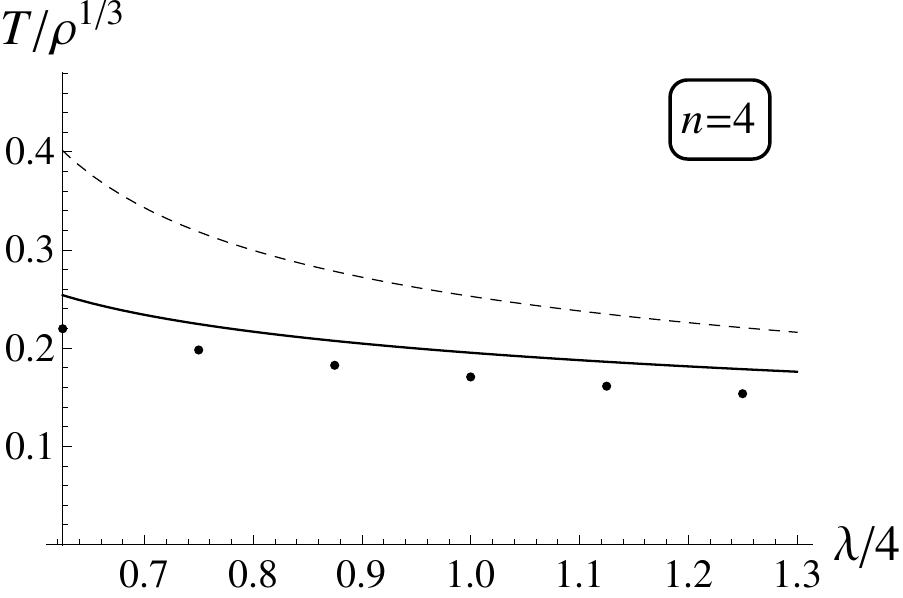}
\qquad
\includegraphics[width=.46\textwidth]{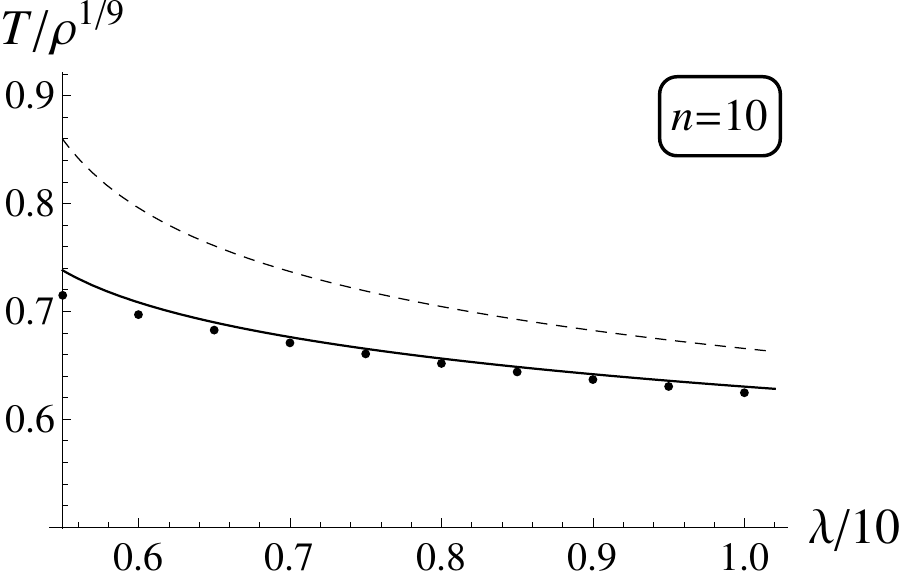}
\end{tabular}
\caption{\small Comparison between calculations of $T/\rho^{1/(n-1)}$ at the critical point. Dots: numerical solution of the exact equations. Solid lines: analytical result \eqref{critT1} (with $\hat\mu=\hat\mu_\mathrm{crit}$). Dashed lines: `classical' value \eqref{critT0} (with $\hat\mu=\hat\mu_c$). The top graphs represent the critical parameter as a function of the boundary spacetime dimension $n$, for fixed values values of the conformal dimension $\lambda$. The bottom graphs represent it as a function of $\lambda$ for fixed values of $n$.
}\label{fig:compare}
\end{figure}

In figure~\ref{fig:compare} we test the accuracy of this result against values that we have computed by numerical integration of eqs.~\eqref{scaleq}, \eqref{gaugeeq}. In appendix~\ref{app:tables} we tabulate more results that we have obtained. It is apparent from these data that \eqref{critT1} determines the critical point with good precision even at low, `realistic' values of $n$: better than $15\%$ for $n=4$, and $35\%$ for $n=3$.

For comparison we also include in figure~\ref{fig:compare} the `classical' value \eqref{critT0}. Since $\lambda-n/2=O(n)$, its difference with \eqref{critT1} is small at large $n$, but the correction makes a significant improvement at moderate $n$. Still, it is remarkable that \eqref{critT0}, using $\hat\mu=\hat\mu_c$ which is notably easy to compute, captures quite well the qualitative dependence on both $n$ and $\lambda$, even if the quantitative agreement at low $n$ is, unsurprisingly, not so good.

The large $D$ value \eqref{critT1} becomes more accurate for larger conformal dimension $\lambda$ of the dual operator. The reason is that in this case the scalar is more massive, and therefore more concentrated near the horizon, where large $D$ fields peak strongly. At the opposite end of the mass spectrum, for scalars at the BF bound, $\hat m^2=-1/4$ \ie\ $\lambda =n/2$, our approach breaks down.

\subsection{Remarks}

By assuming that the scalar field is small, the equations in the near-zone have become effectively a linear system. This has made it possible to solve them and determine the critical point. However, this procedure introduces two issues that are worth clarifying.

First, consider the near and far solutions \eqref{nearsolps} and \eqref{farsoln}. When $\hat\mu-\hat\mu_c>O\lp n^{-2/3}\rp$ we have seen that in the overlap zone both are of the form $A_+ \sR^{-\hat\lambda_+}+A_- \sR^{-\hat\lambda_-}$, and matching them requires $(A_+/A_-)_\mathrm{near}=(A_+/A_-)_\mathrm{far}$. It is easy to see from the explicit coefficients that this equation cannot be satisfied. Thus, our construction of the condensate only works when $\hat\mu-\hat\mu_c=O\lp n^{-2/3}\rp$. For larger values of $\hat\mu$ the near-zone analysis requires solving the full non-linear system of equations.

\begin{figure}[t]
\begin{center}
	\includegraphics[width=.45\textwidth]{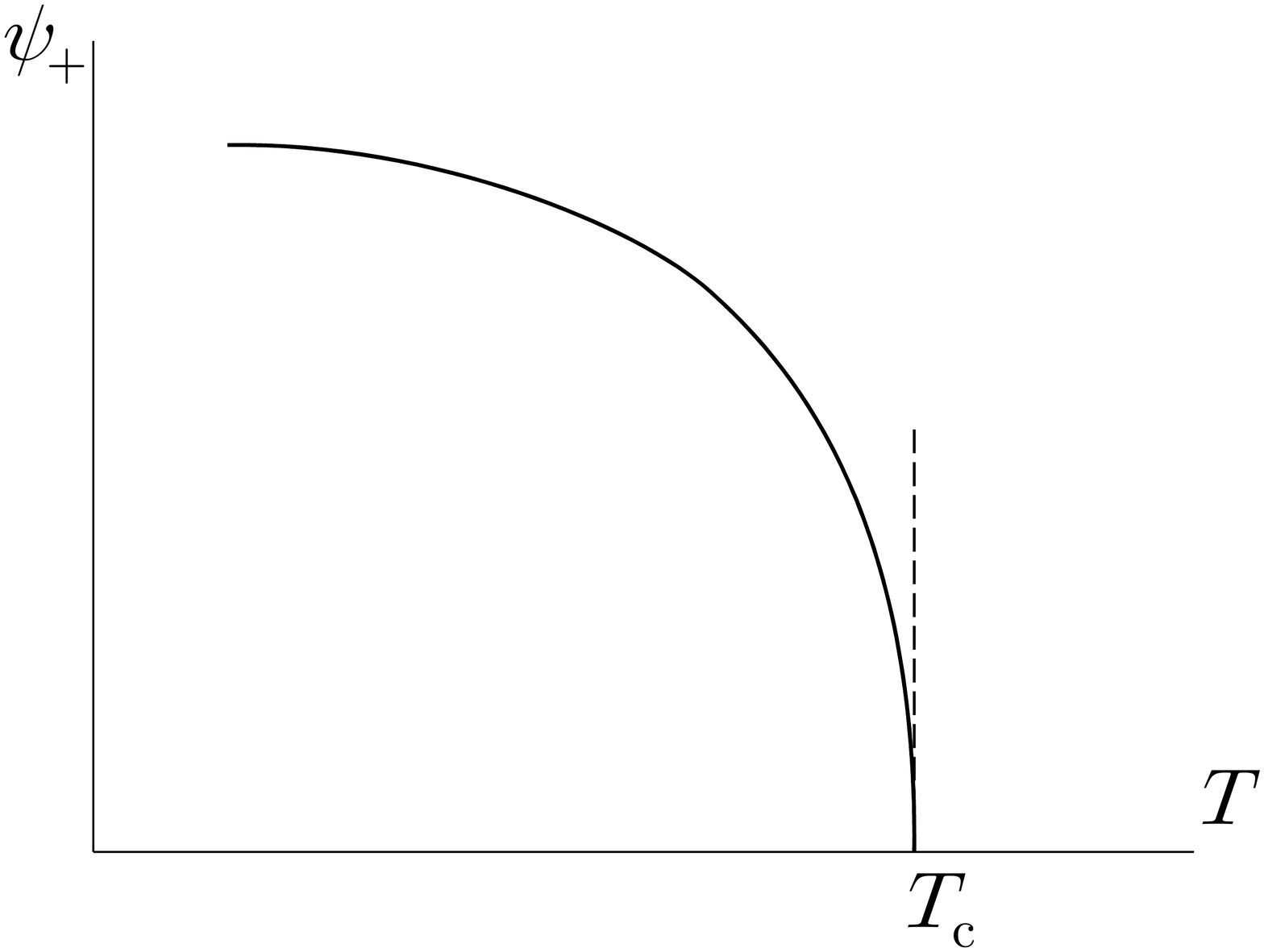}
\end{center}
\caption{\small Solid line: expectation value of the condensate, $\psi_+$, as a function of the temperature in the full non-linear system (the charge density is arbitrarily fixed to $\rho=1$). Dashed line: the condensate in our approximation where $\psi$ solves a linear equation.}
\label{fig:critical}       
\end{figure}

Second, the effective linearization of the system has allowed us to construct a scalar condensate, but its amplitude has been left undetermined, while the control parameter $\hat\mu$ is fixed to the critical value. This appears at odds with the expected behavior of the full non-linear system: there, the amplitude of the condensate is fixed for any value of the control parameter $\hat\mu$, which we are free to vary, and at $\hat\mu=\hat\mu_\mathrm{crit}$ the condensate is not just small: it is zero. The explanation for these differences is that our simplification near the critical point effectively erases all the non-linear information required to move away from it. This is illustrated in fig.~\ref{fig:critical}. As a consequence, we cannot determine the critical exponent at this level of approximation. 
Another limitation is that we cannot obtain the conductivity in the superconducting phase: when we neglect the effect of the scalar on the gauge field, the latter is in the normal, ungapped phase.

While the $SL(2,\R)$ symmetry of the geometry near the horizon may help for going beyond our current approximation and solving the non-linear system, we will not pursue it here.

\section{Simplified analysis and universal features at large $D$}
\label{sec:simple}

One can discern from the previous analysis that establishing the existence and properties of the critical point requires very little information --- and most of it can be extracted from the system in the overlap zone. In particular the only aspect of the near-zone solution \eqref{nearsolps} that we use is that the horizon allows (and requires) that the term $\sR^{-1/2}\ln\sR$ be present in the overlap zone. Recall also that the BF-type bound associated to the onset of the instability is determined by the field in  the overlap zone. This can be reached from the far zone, where the background geometry is that of AdS without a black brane. 

Thus it would seem that a simple study of linear test scalar fields in AdS should yield our result \eqref{critT1} for the critical point. The following analysis shows this more clearly.

\subsection{Scalar condensate as a bound state}
\label{sec:bound}

We introduce a new radial variable
\beq
dr_*=\frac{dr}{h}
\eeq
which is similar to but not the same as the conventional tortoise coordinate for \eqref{adsbb}. The explicit form of $r_*$ as a hypergeometric function of $r$ can be found in \cite{Emparan:2013moa}. We only need to know that $r_*(r)$ is a monotonic function such that
\beqa
r_* &\to& r \quad \mathrm{in~the~far~zone}\,,\\
r_* &\to & r_0\lp 1+\frac{1}{n}\ln (\sR-1)\rp\quad \mathrm{in~the~near~zone}\,,
\eeqa
and in particular $r_*\to -\infty$ at the horizon.
We also redefine the scalar field to
\beq
\chi(r_*)=e^{\frac{n+1}{2}\int \frac{dr}{r}h}\;\psi\,.
\eeq
Then \eqref{scaleq1} becomes
\beq\label{etaeq}
\frac{d^2\chi}{dr_*^2}-\frac{n^2}{r_0^2} V(r_*) \chi=0\,,
\eeq
where
\beq\label{Vrst}
V(r_*)=\frac{r_0^2}{r^2}\left[ 
\frac12+h \lp h\frac{r_0^2}{r^2}\, \hat\mu^2-\hat\mu_c^2-\frac12\rp
\right] \,,
\eeq
after discarding $O(1/n)$ terms and trading in $\hat m$ for $\hat\mu_c$ in \eqref{muc}.
In this guise, the problem of finding what values of $\hat\mu$ allow the scalar condensate to form becomes the problem of tuning $\hat\mu$ in the potential $V$ in such a way that it admits a zero-energy bound state.

It is not difficult to find the overall features of this potential. Asymptotically,  $V\to 0$ at infinity, while $V\to 1/2$ at the horizon. In order to find a bound state, $V$ must have a minimum. If $\hat\mu^2 < \hat\mu_c^2/2$ there is no minimum, but for larger values of $\hat\mu$ there is one at
\beq
r=r_\mathrm{min}=r_0\lp \frac{n}{4}\frac{1+4\hat\mu^2-2\hat\mu_c^2}{2\hat\mu^2-\hat\mu_c^2}+O(n^0)\rp^{1/n}\,.
\eeq
This precise value will not matter much, only that this corresponds to 
\beq
r_{*\mathrm{min}}= r_0\lp 1+O(n^{-1}\ln n)\rp
\eeq
and that the potential at this minimum is
\beq
V_0= \hat\mu_c^2-\hat\mu^2+O(1/n)\,. 
\eeq
It is clear that a zero-energy bound state is impossible if $\hat\mu<\hat\mu_c$. When $\hat\mu=\hat\mu_c$ a `classical' zero-energy state sits at this minimum, but the true bound state requires that $V_0$ dips below zero.

\begin{figure}[t]
\begin{center}
	\includegraphics[width=.7\textwidth]{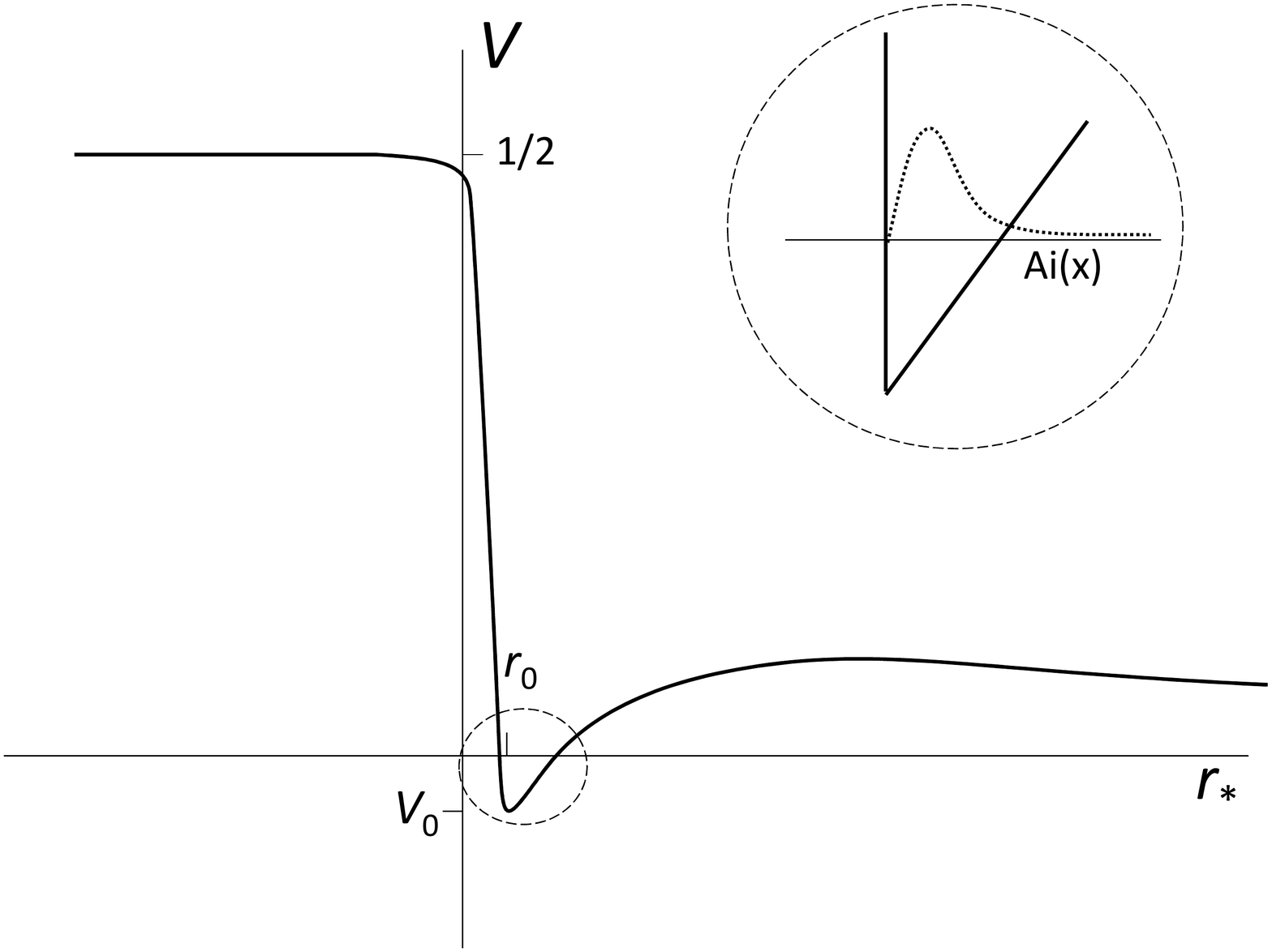}
\end{center}
\caption{\small The potential $V(r_*)$ in \eqref{etaeq} at large $n$ for $\hat\mu>\hat\mu_c$. When $n\to\infty$ the minimum becomes sharp and approaches a triangular potential (inset). When $\hat\mu=\hat\mu_\mathrm{crit}$, a zero-energy bound state exists, with Airy wavefunction Ai$(x)$ (dotted). This is the scalar condensate.}
\label{fig:potential}       
\end{figure}

$r_{*\mathrm{min}}$ lies right in the overlap zone. To its right, there is the far zone where the terms $(r_0/r)^n$ are exponentially small in $n$. To its left, the potential rises exponentially fast in $n$ to $V\to 1/2$ (see fig.~\ref{fig:potential}). Thus at $n\to\infty$ the potential becomes
\beq
V(r_*)= \frac{\Theta(r_0-r_*)}{2}+\Theta(r_*-r_0)V^\mathrm{far}(r_*)
\eeq
where 
\beq\label{Vfar}
V^\mathrm{far}(r)=
\frac{r_0^2}{r^2}\lp
\frac{r_0^2}{r^2}\hat \mu^2-\hat\mu_c^2\rp
\eeq
is the potential \eqref{Vrst} when neglecting the presence of the black hole, $h\to 1$.

It is now very simple to find the lowest value of $\hat\mu$ for which this potential admits a zero-energy bound state. Expanding around the tip,
\beq
V^\mathrm{far}(r_*)=V_0+ (r_*-r_0)V'_0  +O(r_*-r_0)^2\,,\qquad V'_0=\frac{2}{r_0}\lp\hat\mu_c^2-2\hat \mu^2\rp\,,
\eeq
we have a triangular potential, whose bound states are well known to be Airy functions
\beq
\chi(r_*) = \mathrm{Ai} \lp \frac{n^{2/3}}{{V_0'}^{2/3}}\lp V_0+(r_*-r_0) V'_0 \rp\rp\,.
\eeq
If the argument of this function is to remain finite as $n\to\infty$, then $|V_0|$ must approach zero in that limit. This implies that the potential $V$ grows at $r_*=r_0$ infinitely larger than $|V_0|$, and therefore we must impose $\chi(r_*=r_0)=0$. So the required bound state forms when
\beq\label{bounda0}
\frac{V_0}{{V_0'}^{2/3}}=-\frac{a_1}{n^{2/3}}
\eeq
where $-a_1$ is the first Airy zero \eqref{azero}. It is straightforward to see that this reproduces precisely our previous result \eqref{mucrit}.

\subsection{Universal analysis of critical points at large $D$}

Essentially all the information for deriving the bound state comes from the far-region potential $V^\mathrm{far}$. The only input that is required from the horizon is of a qualitative kind: the presence of the black hole induces large gradients that cut-off the potential $V^\mathrm{far}$ at $r_*=r_0$, where it jumps abruptly from a small minimum to a much larger value.

Let us see how this analysis ties in with our previous discussion of BF bounds. For $r_*>r_0$ we have 
\beq
\chi=r^{-(n+1)/2}\psi
\eeq
which satisfies the equation
\beq
\frac{d^2\chi}{dr^2}-\frac{n^2}{r^2}\lp
\frac{r_0^2}{r^2}\hat \mu^2-\hat\mu_c^2\rp \chi=0\,.
\eeq
Near the AdS boundary, at large $r$, the absence of tachyonic behavior requires $\hat\mu_c^2\geq 0$: this is the AdS BF bound. 

Near $r=r_0$, we see that a negative minimum of the potential (hence a possible bound state) appears iff $\hat\mu^2>\hat\mu_c^2$. This is the violation of the near-zone BF bound. The condition \eqref{bounda0} for the formation of the bound state then follows as explained after \eqref{Vfar}.

This analysis can be immediately extended to many other systems of holographic superconductivity as follows:
\begin{itemize}
\item Introduce a large $n$ scaling that allows to neglect the effect of the scalar field on the gauge field, \eg\ $|\Psi|/|A_\mu|\sim 1/n^a$ ($a>0$). This sets the focus on the critical point.
\item In the equation for the scalar field, neglect all terms that fall-off exponentially fast in $n$. This far-zone limit implies that: the gauge field enters the equation only as a constant potential; in the background, the curvature created by the black hole is neglected. The black hole is replaced by the boundary condition that the scalar field vanishes at $r=r_0$  --- physically, $r_0$ is associated to the black hole temperature.
\item If the scalar equation is a linear one (\ie\ the scalar potential $U(\Psi)$ in the Lagrangian is quadratic), then this is a quantum-mechanical problem for 
\beq
\chi=(\sqrt{-g}g^{rr})^{1/2}\psi
\eeq
of the form
\beq
\frac{d^2\chi}{dr^2}-n^2 V(r)\chi=0\,,
\eeq
where $g_{\mu\nu}$ and the potential $V(r)$ are obtained from the system in the background \textit{without} the black hole. If the system's parameters are tuned such that, at $r=r_0$, $V_0<0$ and $V'_0>0$ satisfy \eqref{bounda0}, then the resulting triangular potential admits a zero-energy bound-state condensate. If also asymptotic stability at $r\to\infty$ is maintained then this is the superconducting condensate.
\end{itemize}

This gives a way of determining superconducting transitions in terms of just two numbers $V_0$ and $V_0'$ obtained from the potential in the background geometry at $r=r_0$. For instance, this can be applied to black holes in global AdS. When the scalar potential is not simply quadratic, the problem is not directly one of linear quantum mechanics. It may still be possible to reduce it to a linear problem if one assumes that the condensate scales like a negative power of $n$ (or the parameters in the potential scale with $n$ appropriately), so that higher powers of $\Psi$ are suppressed. 

\section{Including backreaction}
\label{sec:backr}

So far we have considered two limits that simplify the study of holographic superconductors: first, the charge $q$ of the scalar field is very large so we neglect the backreaction of the gauge and scalar fields on the geometry. Then we have taken large $n$, with the gauge and scalar amplitudes being $\phi\sim n$, $\psi\sim n^a$, $a<1$. We have implicitly assumed that $n/q\to 0$, so that the gauge field backreaction is negligible. If instead we keep $n/q$ finite as $n\to\infty$, then this backreaction must be included. However, we can still neglect the backreaction of the scalar field, so the system remains within the remit of the general analysis of the previous section.

In more detail, the starting action is
\beq
I=\int d^{n+1}x\sqrt{-g}\left[ R+n(n-1)-\lp \frac14 F^2 +\left| \nabla\Psi-i q A\Psi\right|^2+m^2|\Psi|^2 \rp\right]\,.
\eeq
The backreaction of the Abelian Higgs sector on the geometry is suppressed when $q\to\infty$ keeping $q A_\mu$ and $q \Psi$ finite. Now, instead, we regard $n$ as a large parameter and keep finite
\beq
\hat q=\frac{q}{n}\,,\qquad \hat m=\frac{m}{n}
\eeq
and
\beq
\hat A_\mu = \hat q A_\mu\,,\qquad 
\hat\Psi = n\hat q\Psi\,.
\eeq
In the large $n$ counting, derivatives count as $\nabla \sim n$. The action becomes a sum
\beq
I=I_{n^2}+I_{n^0}
\eeq
where the $O(n^2)$ part is the metric and gauge field sector,
\beq\label{emads}
I_{n^2}=\int d^{n+1}x\sqrt{-g}\left( R+n(n-1)-\frac1{4\hat q^2} \hat F^2 \right)
\eeq
and the  $O(n^0)$ part is the scalar sector, 
\beq
I_{n^0}=-\frac1{\hat q^2}\int d^{n+1}x\sqrt{-g}\left(\left| \frac1{n}\nabla\hat\Psi-i \hat A\hat\Psi\right|^2 +\hat m^2|\hat\Psi|^2 \right)\,,
\eeq
whose backreaction on the previous sector is therefore negligible at large $n$. 

The background is now a solution of the Einstein-Maxwell-AdS theory \eqref{emads}, which we take to be the Reissner-Nordstrom-AdS black brane. It takes the form of \eqref{adsbb} with
\beq\label{rnh}
h(r)=1-\frac1{r^{n}}\lp r_0^n+\frac{\varrho^2}{2r_0^{n-2}}\rp +\frac{\varrho^2}{2 r^{2n-2}}\,,
\eeq
with outer horizon at $r=r_0$, and gauge field
\beq
\hat\phi(r)=\sqrt{\frac{n-1}{n-2}}\frac{\hat q\varrho}{r_0^{n-2}}\lp 1-\lp\frac{r_0}{r}\rp^{n-2}\rp\,.
\eeq
When $n\to\infty$ the charge density is 
\beq
\rho =q \varrho = n\hat q\varrho\,.
\eeq
The temperature is
\beq
T=\frac{nr_0}{4\pi}\lp 1-\frac{n-2}{n}\frac{\varrho^2}{2 r_0^{2(n-1)}}\rp\,.
\eeq
In order to make contact with our previous analysis it is convenient to introduce
\beq
\hat\mu=\frac{\rho}{n r_0^{n-1}}=\frac{\hat q\varrho}{r_0^{n-1}}\,.
\eeq
Then, at large $n$, \eqref{Trhomu} is replaced by
\beq\label{Trhomu2}
\frac{T}{\rho^{1/(n-1)}}=\frac{n}{4\pi}(n\hat\mu)^{-\frac1{n-1}}\lp1-\frac{\hat\mu^2}{2\hat q^2}\rp\,.
\eeq
The `test gauge field' approximation is recovered when $\hat q$ is large with $\hat q\varrho$ fixed. 

In order to find the scalar condensate we apply the ideas of the previous section. The scalar field equation is the same as \eqref{etaeq}, \eqref{Vrst}, with $\hat\mu_c$ related to the conformal dimension like in \eqref{lambdamuc}, but with $h$ given by \eqref{rnh}. This difference is immaterial, since we are instructed to disregard terms that for $r>r_0$ fall off exponentially fast in $n$ and thus $h\to 1$. The bound state problem is exactly the same as before, and the critical value of $\hat\mu$ is given by the same function $\hat\mu_\mathrm{crit}$ of the conformal dimension as \eqref{mucrit}, which is independent of the backreaction parameter $\hat q$. 

Thus all the dependence of the critical temperature on $\hat q$ is visible in \eqref{Trhomu2}: it is suppressed relative to the value without backreaction, owing  solely to the reduction of the black hole temperature by the charge. This was indeed observed numerically in \cite{Hartnoll:2008kx}. 

\begin{figure}[t]
\begin{center}
\begin{picture}(0,0)(0,0)
\put(-55,70){$\lp T/\sqrt{\varrho}\rp_\mathrm{crit}$}
\put(115,-10){$q$}
\end{picture}
  \includegraphics[width=.5\textwidth]{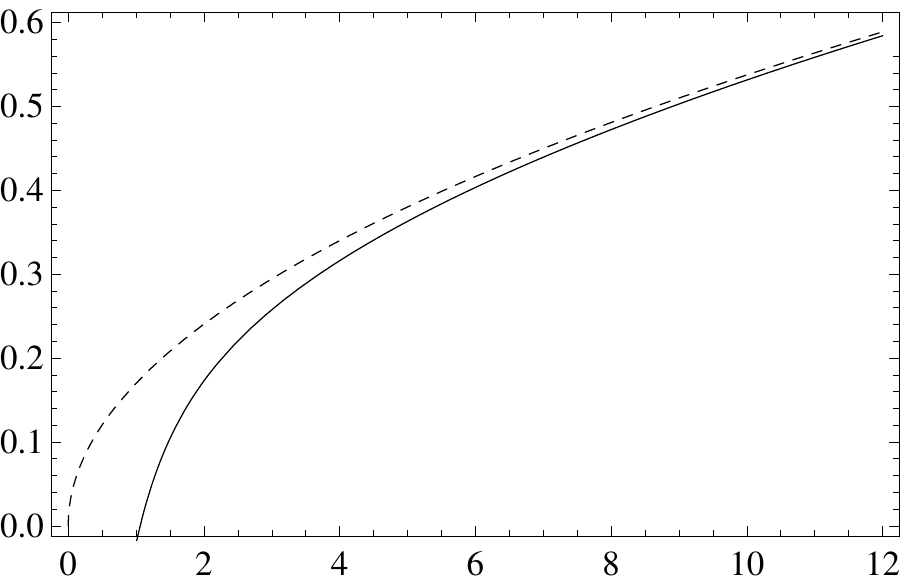}
\end{center}
\caption{\small Critical temperature as a function of charge $q$, for $n=3$, $\lambda=2$. Solid: with gauge field backreaction. Dashed: test gauge field result $\lp T/\sqrt{\varrho}\rp_\mathrm{crit}=0.170\sqrt{q}$. To compare these analytical results with numerical values, see fig.~2b of \cite{Hartnoll:2008kx}.}
\label{fig:backr}       
\end{figure}
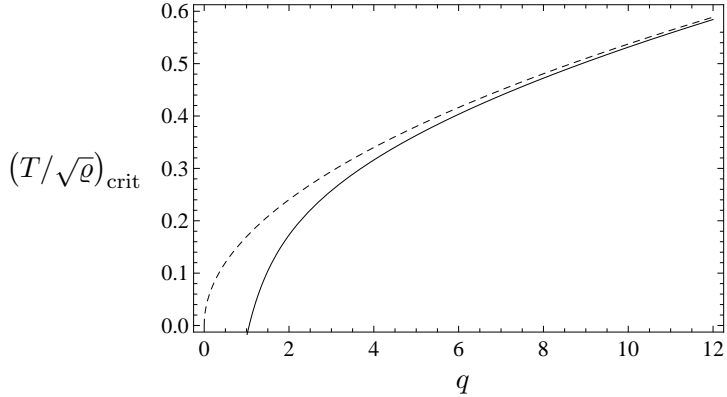

In fig.~\ref{fig:backr} we represent the effect of the backreaction in $T/\sqrt{\varrho}|_\mathrm{crit}=\sqrt{q}\,T/\sqrt{\rho}|_\mathrm{crit}$ for $n=3$, $\lambda =2$. This must be compared with the numerical calculations presented in fig.~2b of \cite{Hartnoll:2008kx}. The agreement is remarkably good for such a low value of $n$. Observe however the discrepancy at very low temperatures: in our result \eqref{Trhomu2}, the critical temperature vanishes for the extremal solution with $\hat q=\hat\mu_\mathrm{crit}/\sqrt{2}$, whereas \cite{Hartnoll:2008kx} find that the condensate persists at lower values of the charge. The explanation of this phenomenon is that as extremality is approached, a throat approximating AdS$_2$ forms very near the horizon, and the violation of the corresponding BF bound can trigger an instability even for neutral fields \cite{Denef:2009tp,Horowitz:2009ij,Hartnoll:2011fn} --- similarly to the mechanism explained in sec.~\ref{sec:nearsol}, but here only applying very close to extremality. The simplified approach of sec.~\ref{sec:simple} cannot capture this effect, since it does not see any of the finite-$\sR$ structure near the horizon. A brief study in appendix~\ref{app:nearRN} of the near-zone along the lines of sec.~\ref{sec:nearsol} reveals that indeed the approach breaks down in the extremal limit, and that slightly higher critical temperatures may be expected close to extremality.

At any rate, we find remarkable that, away from extremality, the large $D$ approach yields the backreacted critical point almost effortlessly. 

\section{Outlook}
\label{sec:concl}

We have shown that the large $D$ expansion affords efficient analytical computation of the critical point of holographic superconductors. Once the basics of the problem are understood, it is reduced to a familiar bound-state calculation. On the downside, the method employs an effective linearization of the system that focuses strongly on the critical point and misses non-linear information such as critical exponents and conductivity in the gapped phase.

The quantitative accuracy that we have achieved without going to high orders in the expansion is encouraging, but some may remain underwhelmed by it: if the point is to get numerically accurate values, one can readily compute them in any specific case by standard numerical integration of the equations (as indeed we have done). 
Probably more significant is that we have learned that holographic superconductors are systems with simple, generic features at large $D$: 
\begin{itemize}
\item The superconducting instability is localized near the horizon in the form of a violation of a BF-type bound.
\item The scalar condensate at the critical point is a zero-energy bound state in a triangular potential. As a consequence the critical temperature is determined simply by the value and the slope of the potential at the minimum.
\end{itemize}
We emphasize that these observations are not merely qualitative but in fact quantitatively sharp.

There have been previous analytic calculations of the critical point based on other approaches. For instance, ref.~\cite{Hartnoll:2008kx} estimates it by a variational approach; ref.~\cite{Herzog:2010vz} modifies the original model in such a way that it can be solved analytically; ref.~\cite{Gregory:2009fj} obtains a simple solution through an uncontrolled approximation (but with one free `matching-point' parameter that can be used to fit the data). The large $D$ expansion provides a perspective on the problem which is complementary to these methods. 

Another boundary-value black hole problem that is solved in a very similar way, also exhibiting universal properties, is that of quasinormal modes, which will be described elsewhere. Given its success with these problems and with earlier studies (see \cite{Hod:2011zzb,Giribet:2013wia,Prester:2013gxa} for other recent applications), we expect that the large $D$ expansion will be of further use in black hole physics.

\section*{Acknowledgments}

RE thanks Antonio Garc{\'\i}a-Garc{\'\i}a and Jan Zaanen for making him aware of the literature on the large $D$ limit in condensed matter theory (even if no use is made of these ideas in this article), and is grateful for hospitality at the Isaac Newton Institute, Cambridge during the workshop ``Mathematics and Physics of the Holographic Principle''.
We acknowledge support from MEC FPA2010-20807-C02-02, AGAUR 2009-SGR-168 and CPAN CSD2007-00042 Consolider-Ingenio 2010. KT is supported by a grant for research abroad from JSPS.

\addcontentsline{toc}{section}{Appendices}
\appendix

\section{Expansions in overlap zone}

\subsection{}
\label{app:neov}

When $\hat\mu\neq \hat\mu_c$, the near-zone solution \eqref{nearsolps} at large $\sR$ becomes
\beq\label{nearover}
\psi\to \sR^{-\hat\lambda_+}\frac{\Gamma(\hat\lambda_--\hat\lambda_+)}{\Gamma(\hat\lambda_-+i\hat\mu)\Gamma(\hat\lambda_--i\hat\mu)}+
\sR^{-\hat\lambda_-}\frac{\Gamma(\hat\lambda_+-\hat\lambda_-)}{\Gamma(\hat\lambda_++i\hat\mu)\Gamma(\hat\lambda_+-i\hat\mu)}\,.
\eeq
When $\hat\mu=\hat\mu_c$, so that $\hat\lambda_+=\hat\lambda_-=1/2$, one gets instead
\beq
\psi\to \frac{\cosh(\pi\hat\mu_c)}{\pi\sqrt{\sR}}\left(-
2\gamma -\psi(1/2+i\hat\mu_c)-\psi(1/2-i\hat\mu_c) +\ln \sR
\right)\,,
\eeq
with $\gamma$ the Euler constant and $\psi$ the digamma function.

In this last expansion, $\sqrt{R}\,\psi$ has a constant term that is absent from the far zone solution \eqref{farov} at the same order in $n$. However, this is not a difficulty for matching the solutions, as it corresponds to a redefinition between the near and far zone radii, $\sR \to \delta a \sR$. Since it is a gauge issue, the matching of this term does not have physical meaning.

\subsection{}
\label{app:debye}

In the regime in which both the argument and the index of the Bessel function in \eqref{farsoln} are large and of the same order $\sim n$, and with $r$ as in \eqref{overrR}, Debye's asymptotic representation gives
\beq\label{farover1d}
\psi=\frac{1}{\sqrt{2\pi n\hat\mu_c \tanh\alpha}}K\sR^{-\hat\lambda_+}\lp 1+O(1/n)\rp\,, \qquad \hat\mu<\hat\mu_c
\eeq
and
\beq\label{farover2d}
\psi=\frac{1}{\sqrt{2\pi n\hat\mu_c \tan\beta}}\lp K\sR^{-\hat\lambda_+}\lp 1+O(1/n)\rp
+K^{-1}\sR^{-\hat\lambda_-}\lp 1+O(1/n)\rp \rp\,, \quad \hat\mu>\hat\mu_c
\eeq
where
\begin{eqnarray}
K=
\begin{cases}
e^{-n\hat\mu_c(\alpha-\tanh\alpha)}&\quad\hat\mu<\hat\mu_c\,,\\
e^{-in\hat\mu_c(\beta-\tan\beta)-i\pi/4}&\quad\hat\mu>\hat\mu_c\,,
\end{cases}
\end{eqnarray} 
%
and $\alpha$ and $\beta$ defined by
\begin{equation}
\frac{\hat\mu}{\hat\mu_c}=
\begin{cases}
\,\text{sech}\,\alpha\,,&\hat\mu<\hat\mu_c\,,\\
\,\text{sec}\,\beta\,,&\hat\mu_c<\hat\mu\,.
\end{cases}
\end{equation}

\section{Critical points}
\label{app:tables}

Here we tabulate results for the critical temperature $T$ (arbitrarily setting $\rho=1$), comparing the result $T_\mathrm{num}$ from numerical integration of the exact equations, and the analytical result $T_\mathrm{an}$ from eq.~\eqref{critT1}. For ease of interpretation we present the results in two manners: the first more appropriate for fixed $\lambda$, varying $n$; the second, for fixed $n$ and varying $\lambda$.

\begin{table}[h]
\begin{center}
\medskip
\begin{tabular}{|l|c|c||c|c||c|c|}
\cline{2-7}
 \multicolumn{1}{c|}{}&\multicolumn{2}{c||}{$\lambda=\frac{n+1}{2}$}&\multicolumn{2}{c||}{$\lambda=3n/4$}&\multicolumn{2}{c|}{$\lambda=n$}\\
\cline{2-7}
\multicolumn{1}{c|}{} &$T_\mathrm{num}$&$T_\mathrm{an}$&$T_\mathrm{num}$&$T_\mathrm{an}$&$T_\mathrm{num}$&$T_\mathrm{an}$\\
\cline{2-7}
\hline
$n=3$&0.118&0.170&0.108&0.153&0.0867&0.125\\
\hline
$n=4$&0.220&0.254&0.198&0.224&0.171&0.195\\
\hline
$n=5$&0.309&0.336&0.279&0.296&0.249&0.266\\
\hline
$n=6$&0.393&0.417&0.357&0.369&0.326&0.337\\
\hline
$n=7$&0.473&0.497&0.433&0.442&0.400&0.410\\
\hline
$n=8$&0.555&0.578&0.509&0.516&0.475&0.483\\
\hline
$n=9$&0.635&0.658&0.585&0.591&0.550&0.556\\
\hline
$n=10$&0.715&0.734&0.660&0.665&0.624&0.630\\
\hline
\end{tabular}
\end{center}
\end{table}

\begin{table}[h]
\begin{center}
\medskip
\begin{tabular}{|l|c|c||c|c||c|c||c|c|}
\cline{2-9}
\multicolumn{1}{c|}{}&\multicolumn{2}{c||}{$n=3$}&\multicolumn{2}{c||}{$n=4$}&\multicolumn{2}{c||}{$n=9$}&\multicolumn{2}{c|}{$n=10$}\\
\cline{2-9}
\multicolumn{1}{c|}{} &$T_\mathrm{num}$&$T_\mathrm{an}$&$T_\mathrm{num}$&$T_\mathrm{an}$&$T_\mathrm{num}$&$T_\mathrm{an}$&$T_\mathrm{num}$&$T_\mathrm{an}$\\
\cline{2-9}
\hline
$\lambda=n-2$      &-     &-    &-    &-    &0.580&0.585&0.652&0.656\\
\hline
$\lambda=n-\frac32$&-     &-    &0.220&0.254&0.571&0.576&0.644&0.648\\
\hline
$\lambda=n-1$      &0.118 &0.170&0.198&0.224&0.563&0.569&0.637&0.642\\
\hline
$\lambda=n-\frac12$&0.0992&0.141&0.182&0.207&0.556&0.562&0.630&0.636\\
\hline
$\lambda=n$        &0.0867&0.125&0.171&0.195&0.550&0.556&0.624&0.630\\
\hline
$\lambda=n+\frac12$&0.0778&0.115&0.161&0.186&-    &-    &-    &-    \\
\hline
$\lambda=n+1$      &-     &-    &0.153&0.178&-    &-    &-    &-    \\
\hline
\end{tabular}
\end{center}
\end{table}

\newpage

\section{Near-horizon geometry of charged black brane}
\label{app:nearRN}

Introducing the coordinate $\sR$ of \eqref{sR} and defining
\beq
u=\frac{\varrho}{\sqrt{2}\,r_0^{n-1}}=\frac{\hat\mu}{\sqrt{2}\,\hat q}
\eeq
($0\leq u\leq 1$) we obtain the non-trivial two-dimensional part of the large $n$ near-horizon geometry of the charged black brane in sec.~\ref{sec:backr},
\beq\label{nearRN}
ds^2|_\mathrm{nh}= -\lp 1-\frac1{\sR}\rp\lp 1-\frac{u^2}{\sR}\rp d\hat t^2+\frac{d\sR^2}{(\sR-1)(\sR-u^2)}\,.
\eeq
The scalar field equation 
\beq\label{chsceq}
\lp\sR\frac{d}{d\sR}\rp^2\psi+\frac{\sR^2-u^2}{(\sR-1)(\sR-u^2)}\sR\frac{d}{d\sR}\psi
+\frac{\hat\mu^2\sR^2}{(\sR-u^2)^2}\psi -\frac{\hat m^2\sR^2}{(\sR-1)(\sR-u^2)}\psi=0\,.
\eeq
now has four regular singular points, at $\sR=0,\infty,1,u^2$, and therefore it is not obviously of hypergeometric type. But even though it may still admit an explicit solution, we do not need it: by continuity with the case $u=0$, the condition of regularity at the non-extremal horizon is expected to generically imply the presence of the term $\sR^{-1/2}\ln\sR$ in the overlap zone $\sR\to\infty$ when $\hat\mu=\hat\mu_c+O(n^{-2/3})$. This is enough to proceed with the methods of sec.~\ref{sec:bound}.

This argument, however, breaks down when the black brane approaches extremality, $u\to 1$, and the geometry \eqref{nearRN} develops an AdS$_2$ throat. In this limit eq.~\eqref{chsceq} becomes
\beq\label{exsceq}
\lp\sR\frac{d}{d\sR}\rp^2\psi+\frac{\sR+1}{\sR-1}\sR\frac{d}{d\sR}\psi
 -\frac{(\hat m^2-\hat\mu^2)\sR^2}{(\sR-1)^2}\psi=0\,,
\eeq
which is easily solved,
\beq
\psi=A_+(\sR-1)^{-\hat\lambda_+}+A_-(\sR-1)^{-\hat\lambda_-}\,. 
\eeq
This implies that the AdS$_2$ BF bound in the region near $\sR=1$ is the same as in the asymptotic region $\sR\to \infty$. Thus the change in behavior is not due to a different BF bound very close to the horizon. We still expect the transition to happen near $\hat\mu=\hat\mu_c$. But the general solution in this case is
\beq
\psi=\frac{1}{\sqrt{\sR-1}}\lp A+ B \ln\lp\sR-1\rp\rp
\eeq
and regularity at the horizon requires $B=0$. Then in the overlap zone at $\sR\to\infty$ the term $\sR^{-1/2}\ln\sR$ is absent, and the analysis of sec.~\ref{sec:simple} does not apply anymore. The condition for the condensate is modified: instead of \eqref{aicond} we should presumably impose $\mathrm{Ai}'(-\delta\mu)=0$, so $\delta\mu=a_1'=1.0188$ which results in a different, lower value of $\hat\mu_\mathrm{crit}$. Although this does not explain why the condensate may form at arbitrarily low values of $\hat q$, it does suggest that as extremality is approached, we may expect a higher critical temperature than obtained in the approach of sec.~\ref{sec:simple}. It is possible that with further work one can perform a large $D$ construction of the neutral condensate near extremality.


\end{document}